\documentclass
[aps,pra,amsfonts,amssymb,twocolumn,amsmath,preprintnumbers,nofootinbib,floatfix,
showpacs]{revtex4-1}%
\usepackage[dvips]{graphics}
\usepackage{graphicx}
\usepackage{bm}
\usepackage{amsmath}
\usepackage{amsfonts}
\usepackage{amssymb}
\usepackage{xcolor}
\usepackage{subfigure}
\usepackage{hyperref,hypcap}
\usepackage{braket}
\usepackage{commath}%
\providecommand{\U}[1]{\protect\rule{.1in}{.1in}}
\begin{document}

\title{Boltzmann scaling of spontaneous Hall current and nonequilibrium spin-polarization}
\author{Cong Xiao, Bangguo Xiong, Fei Xue}
\affiliation{Department of Physics, The University of Texas at Austin, Austin, Texas 78712, USA}
\begin{abstract}
We extend the semiclassical Boltzmann formalism for the anomalous Hall effect
(AHE) in non-degenerate multiband electron systems to the spin Hall effect
(SHE) and unconventional Edelstein effect (UEE, cannot be accounted for by the
conventional Boltzmann equation, unlike the conventional Edelstein effect).
This extension is confirmed by extending the Kohn-Luttinger density-matrix
transport theory in the weak disorder-potential regime. By performing Kubo
linear response calculations in a prototypical multiband model, the Boltzmann
scaling for the AHE/SHE and UEE is found to be practically valid only if the
disorder-broadening of bands is quite smaller than the minimal intrinsic
energy-scale around the Fermi level. Discussions on this criterion in various
multiband systems are also presented. A qualitative phase diagram is proposed
to show the influences of changing independently the impurity density and
strength of disorder potential on the AHE/SHE and UEE.
\end{abstract}
\maketitle

\section{Introduction}

Disorder effects on nonequilibrium properties of Bloch electrons is a basic
issue in condensed matter physics. Many of them can be discussed within the
relaxation time approximation of the conventional semiclassical Boltzmann
equation \cite{Ziman1960}. However, some transport phenomena related to the
spin-orbit coupling, such as the spin Hall effect (SHE) and anomalous Hall
effect (AHE) \cite{Nagaosa2010,Sinova2015}, contain intriguing
disorder-induced effects that cannot be treated by the conventional Boltzmann
equation \cite{Nagaosa2010,Sinova2015,Sinitsyn2008}. Another
spin-orbit-induced nonequilibrium phenomenon is the Edelstein effect --
nonequilibrium spin-polarization driven by external electric fields
\cite{Edelstein1990}. The conventional Edelstein effect is described by the
conventional Boltzmann equation \cite{Zhang2008}. While the unconventional
Edelstein effect (UEE), in which a nonequilibrium spin-polarization arises in
the direction perpendicular to that in the conventional Edelstein effect
\cite{Garate2009,Taguchi2017}, is related to aforementioned intriguing
disorder effects \cite{Xiao2017SOT-SBE}. In the presence of exchange coupling
to a local magnetization, the conventional and unconventional Edelstein
effects give rise to the fieldlike and dampinglike spin-orbit torques on the
magnetization, respectively \cite{Sinova2015,Xiao2017SOT-SBE}.

Those intriguing effects due to static disorder, including the skew scattering
\cite{Sinitsyn2008}, side-jump
\cite{Sinitsyn2005,Sinitsyn2006,Sinitsyn2008,Xiao2017FOP,Xiao2017SOT-SBE} and
scattering off pairs of impurities \cite{Sinitsyn2007,Xiao2017FOP}, have been
incorporated into the generalized semiclassical Boltzmann formalism by
semiclassical or semi-phenomenological arguments. For the AHE, the generalized
Boltzmann formalism formulated in the weak disorder-potential regime has its
root in the Kohn-Luttinger density-matrix transport approach to electrical
conductivities \cite{Sinitsyn2006,Sinitsyn2008,KL1957,Luttinger1958}. However,
in the case of the SHE and UEE, such a necessary identification is still
absent. In the present paper it will be provided by extending the
Kohn-Luttinger approach to the SHE and UEE \cite{note-SOscattering}.

The Boltzmann formalism yields the Boltzmann scaling \cite{Luttinger1958}
\begin{equation}
\alpha=c_{sk}+\left(  c_{in}+c_{AQ}\right)  \rho_{yy} \label{scaling}%
\end{equation}
in the presence of one type of static disorder. Here $\alpha$ can represent
the anomalous Hall ratio ($\sigma_{xy}/\sigma_{yy}$, $\sigma_{xy}$ is the Hall
conductivity, $\sigma_{yy}$ is the longitudinal conductivity), spin Hall ratio
($\sigma_{xy}^{s}/\sigma_{yy}$, $\sigma_{xy}^{s}$ is the spin Hall
conductivity) and the UEE-efficiency per current (e.g., $\chi_{yy}/\sigma
_{yy}$, $\chi_{yy}$ is the UEE response coefficient). $\rho_{yy}$ stands for
the longitudinal resistivity and $\rho_{yy}\gg\rho_{xy}$ is assumed. $c_{in}$
comes from the intrinsic contribution, whereas $c_{sk}$ and $c_{AQ}$ come from
the skew scattering and anomalous quantum (called side-jump in Refs.
\cite{Nagaosa2010,Sinova2015}) contributions, respectively. These
nomenclatures are explained in Sec. III. A well-defined scaling relation
exists only if the scaling parameters remain constant as the scaling variables
change. In\textit{ }the\textit{ Boltzmann framework} $c_{sk}$, $c_{in}$ and
$c_{AQ}$ remain constant when the impurity density is changed. Thus the
\textit{so tuned} $\rho_{yy}$ plays the role of a scaling variable, and $c$'s
scaling parameters. The multivariable Boltzmann scaling for the AHE in the
presence of more than one type of disorder has also been proposed
\cite{Hou2015} via an approach equivalent to the Boltzmann formalism
\cite{Sinitsyn2007,Kovalev2010,Xiao2017SOT-SBE}. The Boltzmann scaling
(\ref{scaling}) and its multivariable generalization have played the central
role in understanding measurements and analyzing numerical results in the
field of AHE/SHE \cite{Seemann2010,Jin2017,Tian2009,Hou2015,Sagasta2016}.

However, theoretically the regime of validity of Boltzmann scaling remains
unclear. This is the second topic in the present paper. The Boltzmann
formalism is intuitively anticipated to work well only if the
disorder-broadening $\hbar/\tau$\ of bands is quite smaller than the minimal
intrinsic energy scale $\Delta$ of the band structure around the Fermi level.
$\Delta$ is usually the minimal interband splitting around the Fermi level and
depends on the position of the latter. Although some previous researches on
the intrinsic AHE/SHE support this idea
\cite{Kontani2007,Kontani2007Pt,Tanaka2008}, some other work suggest that the
Boltzmann scaling is valid up to $\hbar/\tau\lesssim\epsilon_{F}$
\cite{Onoda2006,Onoda2008} or $\hbar/\tau\lesssim0.1\epsilon_{F}$
\cite{Nagaosa2008}. This situation has caused confusion in understanding
experimental results \cite{Su2014}. Focusing on the case of short-range weak
disorder-potential ($DV_{0}\lesssim0.1$ in practice, $D$ is the typical
density of states around the Fermi level, $V_{0}$ is the Fourier component of
the disorder potential $V\left(  \mathbf{r}\right)  $ at zero wavevector), we
find that the Boltzmann scaling is \textit{practically} or
\textit{approximately} valid if $\frac{\hbar}{\tau}<\frac{\Delta}{\pi}$, i.e.,
$\left(  \frac{\hbar}{\Delta\tau}\right)  ^{2}<0.1$. This is obtained in a
prototypical multiple conduction-band model and found to be applicable in
various other systems. Moreover, a qualitative phase diagram is proposed to
show the influences of changing independently the impurity density and the
strength of disorder potential on the AHE/SHE and UEE.

The present paper is organized as follows. The Boltzmann formulations are
outlined in Sec. II, whereas the regime of validity of the Boltzmann scaling
for AHE/SHE and UEE is analyzed in Sec. III. Section V summarizes the paper.
Appendices A and B include necessary discussions on the semiclassical
Boltzmann formalism, whereas some calculation details are given in Appendix C.

\section{Kohn-Luttinger derivation of the Boltzmann transport}

In the Boltzmann formalism of linear response, the average value of an
observable $A$ (quantum mechanically, Hermitian operator $\hat{A}$) in the
presence of a dc weak uniform electric field $\mathbf{E}$ is given by
$A=\sum_{l}f_{l}A_{l}$, with the index $l$ denoting the carrier state. In the
present paper we consider non-degenerate multiband carrier systems. The
semiclassical distribution function $f_{l}$ is governed by the generalized
semiclassical Boltzmann equation \cite{KL1957,Luttinger1958,Sinitsyn2008} in
nonequilibrium steady-states in the presence of elastic carrier-impurity
scattering. $A_{l}$ can be written as
\cite{Xiao2017SOT-SBE,note-SOscattering}
\begin{equation}
A_{l}=A_{l}^{0}+\delta^{in}A_{l}+\delta^{sj}A_{l}. \label{semi-A-inter}%
\end{equation}
with $\delta^{in}A_{l}=-\hbar e\mathbf{E\cdot}\sum_{l^{\prime}\neq
l}2\mathrm{Im}\langle u_{l}|\mathbf{\hat{v}}|u_{l^{\prime}}\rangle
\delta_{\mathbf{kk}^{\prime}}A_{l^{\prime}l}/d_{ll^{\prime}}^{2}$ and%
\begin{equation}
\delta^{sj}A_{l}=\sum_{l^{\prime},l^{\prime\prime}\neq l^{\prime}}%
\frac{\left\langle V_{ll^{\prime}}V_{l^{\prime\prime}l}\right\rangle
A_{l^{\prime}l^{\prime\prime}}}{d_{ll^{\prime}}^{-}d_{ll^{\prime\prime}}^{+}%
}+2\operatorname{Re}\sum_{l^{\prime}\neq l,l^{\prime\prime}}\frac{\left\langle
V_{l^{\prime}l^{\prime\prime}}V_{l^{\prime\prime}l}\right\rangle
A_{ll^{\prime}}}{d_{ll^{\prime}}^{+}d_{ll^{\prime\prime}}^{+}}. \label{ex}%
\end{equation}
Here $|l\rangle=|\mathbf{k}\rangle|u_{l}\rangle$ is the Bloch state,
$l=\left(  \eta,\mathbf{k}\right)  $ with $\eta$ the band index and
$\mathbf{k}$ the momentum. $\left\langle ..\right\rangle $ represents the
average over disorder configurations, $d_{ll^{\prime}}\equiv\epsilon
_{l}-\epsilon_{l^{\prime}}$, $d_{ll^{\prime}}^{\pm}\equiv d_{ll^{\prime}}\pm
i\hbar s$ with $s\rightarrow0^{+}$. In the case of $\hat{A}=\mathbf{\hat{v}}$,
$\delta^{in}\mathbf{v}_{l}$ and $\delta^{sj}\mathbf{v}_{l}$ coincide with the
Berry-curvature anomalous velocity \cite{Sinitsyn2008} and the side-jump
velocity \cite{Xiao2017SOT-SBE}, respectively. Both of them have microscopic
derivations \cite{Luttinger1958}. While, in other cases $\delta^{in}A_{l}$ and
$\delta^{sj}A_{l}$ were only added into the Boltzmann formalism
semi-phenomenologically \cite{Xiao2017SOT-SBE}.

In the present paper we give the microscopic derivation to Eq.
(\ref{semi-A-inter}) in the case of $\hat{A}$ other than $\mathbf{\hat{v}}$.
Because this justification is obtained by resorting to the Kohn-Luttinger
density-matrix approach \cite{KL1957}, we provide it in Appendix A in order
not to introduce too many notations in the main text. From that derivation one
can see that, Eq. (\ref{semi-A-inter}) accounts for the off-diagonal response
of the out-of-equilibrium single-particle density-matrix in the
band-eigenstate representation \cite{Culcer2017}.

\section{Regime of validity of Boltzmann Scaling}

\subsection{Two-conduction-band model calculation}

We consider the 2D Hamiltonian
\begin{equation}
\hat{H}_{0}=\frac{\hbar^{2}\mathbf{k}^{2}}{2m}+\alpha_{R}\mathbf{\hat{\sigma}%
}\cdot\left(  \mathbf{k}\times\mathbf{\hat{z}}\right)  -\epsilon_{I}%
\hat{\sigma}_{z}, \label{model}%
\end{equation}
where $m$ is the effective mass of conduction electron, $\mathbf{k}=k\left(
\cos\phi,\sin\phi\right)  $ the 2D wavevector, $\mathbf{\hat{\sigma}}=\left(
\hat{\sigma}_{x},\hat{\sigma}_{y},\hat{\sigma}_{z}\right)  $ are the Pauli
matrices for\ electron spin. In different qualitative realizations of this
Hamiltonian, $\alpha_{R}>0$ and $\epsilon_{I}>0$ have different physical
interpretations. In ultrathin ferromagnets embedded between two asymmetric
interfaces \cite{Li2015}, $\alpha_{R}$ is the Rashba spin-orbit coupling
coefficient, $\epsilon_{I}$ is the exchange coupling. In gated
transition-metal dichalcogenides \cite{Law2014,Taguchi2017,Zhou2017},
$\alpha_{R}$ describes the Rashba coupling due to the gating field,
$\epsilon_{I}$ refers to the Ising spin-orbit coupling arising from in-plane
mirror symmetry breaking, and the Ising term takes opposite values ($\pm$) in
opposite valleys. More importantly, this Hamiltonian serves as a minimal model
for multiband systems with multiple-Fermi-surfaces and avoided
band-anticrossing point \cite{Nagaosa2010}. In this case $\epsilon_{I}$ plays
the role of the spin-orbit coupling that lifts the accidental degeneracy of
band dispersions with the velocity $\alpha_{R}/\hbar$
\cite{Onoda2006,Onoda2008}. Although this special 2D model breaks both the
inversion and time reversal symmetries, some generic qualitative insights can
still be acquired, which apply to various ferromagnetic
\cite{Nagaosa2010,Onoda2006,Onoda2008,Miyasato2007} and nonmagnetic
\cite{Hoffmann2013} materials possessing multiple-Fermi-surfaces.

For any energy $\epsilon>\epsilon_{I}$ there are two iso-energy rings
corresponding to the two subbands $\eta=\pm$: $k_{\eta}^{2}\left(
\epsilon\right)  =\frac{2m}{\hbar^{2}}\left(  \epsilon-\eta\Delta_{\eta
}\left(  \epsilon\right)  \right)  $ where $\epsilon_{R}=m\left(  \frac
{\alpha_{R}}{\hbar}\right)  ^{2}$ and $\Delta_{\eta}\left(  \epsilon\right)
=\sqrt{\epsilon_{I}^{2}+\epsilon_{R}^{2}+2\epsilon_{R}\epsilon}-\eta
\epsilon_{R}$. We focus on the case where both subbands are partially occupied
in our analytic treatment, whereas the regime $\epsilon_{F}<\epsilon_{I}$ will
also be addressed later (Sec. III. B). The qualitative insights obtained in
the former case can also be applied to the latter one. Randomly distributed
identical $\delta$-scatterers are assumed. For the simplest assumption of
scalar disorder, the pathological properties of model (\ref{model}) in the
case of both subbands partially occupied, e.g., the vanishing AHE/SHE and UEE
irrespective of the impurity density under the noncrossing approximation in
the case of weak disorder potential
\cite{Nagaosa2010,Sinova2015,Xiao2017SOT-SBE,Xiao2017SOT}, make it
inconvenient to extract general insights. Fortunately, one can get around this
inconvenience by just assuming another type of short-range disorder $\hat
{V}=V_{0}\hat{\sigma}_{z}$. Although such a kind of disorder has its root in
realistic considerations as detailed in Refs.
\cite{Yang2011,Pesin2012,Xiao2017SOT-SBE}, we just regard it as an approach
that gets around the pathological properties of the Rashba model and makes the
model a prototypical one from which general \textit{qualitative} insights can
be extracted. In the following we will focus on the UEE in this model, whereas
the considerations on the AHE and SHE are completely similar. We will only
focus on the aspects of the model that can be meaningful for general multiband systems.

\subsubsection{\textbf{Boltzmann calculation}}

In model (\ref{model}), due to the UEE there is a nonequilibrium spin density
$\left(  \delta\mathbf{S}\right)  _{||}$ parallel to the driving electric
field. The Boltzmann theory yields $\left(  \delta\mathbf{S}\right)
_{\parallel}=\delta^{in}\mathbf{S}+\delta^{AQ}\mathbf{S}+\delta^{sk}%
\mathbf{S}$. Here $\delta^{in}\mathbf{S}$ is the intrinsic contribution,
$\delta^{AQ}\mathbf{S}$ is termed the anomalous quantum contribution which
arises from disorder but turns out to be independent of the impurity density,
$\delta^{sk}\mathbf{S}$ is the skew scattering contribution inversely
proportional to the impurity density. The anomalous/spin Hall current
\textit{within the Boltzmann framework} can also be parsed in the same way.
Two necessary notes on the Boltzmann calculation are in order.

First, in the weak disorder-potential regime $\delta^{sk}\mathbf{S}$ is
dominated by the contribution from $o\left(  V^{3}\right)  $ non-Gaussian
disorder correlation $\left\langle V^{3}\right\rangle _{c}\sim n_{im}V_{0}%
^{3}$. Here $n_{im}$ is the impurity density, $\left\langle ..\right\rangle
_{c}$ is the connected part of disorder correlation. The transport time of the
$o\left(  V^{3}\right)  $ skew scattering is of scale $DV_{0}\tau$ (Appendix
C). The higher-order skew scattering is usually negligible compared to the
$o\left(  V^{3}\right)  $ one in the weak disorder-potential regime, because,
e.g., the transport time of the $o\left(  V^{4}\right)  $ skew scattering
\cite{Kovalev2008,Kovalev2009,Xiao2017AHE} is of scale $\left(  DV_{0}\right)
^{2}\tau\ll DV_{0}\tau$ (Appendix C).

Second, the effect of scattering off pairs of impurities enters into
$\delta^{AQ}\mathbf{S}$ via both the noncrossing-diagram and crossing-diagram
parts of $\omega_{ll^{\prime}}$ in $o\left(  V^{4}\right)  $ ($\omega
_{ll^{\prime}}$ is the semiclassical scattering rate
\cite{Xiao2017SOT-SBE,Sinitsyn2008}, see Appendix B). We only address the
noncrossing-diagram part in the concrete calculation. For the purpose of this
paper, the quantitative difference due to the inclusion of the crossing part
\cite{Ado} is unimportant.

Concrete calculations of $\delta S_{y}=$ $\chi_{yy}E_{y}$ have been presented
in Appendix C and Ref. \cite{Xiao2017SOT}. Here we write down the UEE
efficiency $\alpha=\chi_{yy}/\sigma_{yy}$ in a form%
\begin{align}
\alpha &  =-e\alpha_{R}D_{0}\left[  \frac{1}{\frac{e^{2}\epsilon_{F}}{\pi
\hbar^{2}}}\frac{mV_{0}}{\hbar^{2}}\frac{f_{sk}^{UEE}\left(  C_{1},0\right)
}{f_{L}\left(  C_{1},0\right)  }\right. \nonumber\\
&  \left.  +f_{in+AQ}\left(  C_{1},D_{2},0\right)  \rho_{yy}\right]  ,
\label{scaling-UEE-SB}%
\end{align}
which is convenient to be compared with the corresponding result Eq.
(\ref{scaling-UEE}) obtained in the Kubo-Streda formula. Here $D_{0}=\frac
{m}{2\pi\hbar^{2}}$, and other notations are described in the next subsection.
What is important is that $f_{sk}^{UEE}\left(  C_{1},0\right)  $,
$f_{L}\left(  C_{1},0\right)  $ and $f_{in+AQ}\left(  C_{1},D_{2},0\right)  $
are all independent of both the impurity density and disorder potential. Thus
when tuning $\rho_{yy}$ via changing the impurity density, Eq.
(\ref{scaling-UEE-SB}) is just the Boltzmann scaling (\ref{scaling}).

\subsubsection{\textbf{Kubo calculation}}

The linear response to a dc uniform electric field in the single-particle
picture with static disorder can be found by the Kubo-Streda formula
\cite{Streda1977}. In the weak disorder-potential regime not far away from the
weak disorder-potential limit, one can apply the standard ladder approximation
and consider the conventional Mercedes star diagrams for the $o\left(
V^{3}\right)  $\ skew scattering \cite{Borunda2007}, leading to $\chi
_{yy}=\chi_{yy}^{b}+\chi_{yy}^{l}+\chi_{yy}^{sk}$ at the zero-temperature
limit with
\begin{align*}
\chi_{yy}^{b}  &  =-e\alpha_{R}D_{0}\tau iI_{2},\\
\chi_{yy}^{l}  &  =-e\alpha_{R}D_{0}f_{in+AQ}\left(  I_{1},\tau iI_{2}%
,I_{2}^{2}\right)  ,\\
\chi_{yy}^{sk}  &  =-e\alpha_{R}D_{0}\tau\frac{mV_{0}}{\hbar^{2}}f_{sk}%
^{UEE}\left(  I_{1},I_{2}^{2}\right)  ,
\end{align*}
and $\sigma_{yy}=\frac{e^{2}}{\pi\hbar}\frac{\epsilon_{F}\tau}{\hbar}%
f_{L}\left(  I_{1},I_{2}^{2}\right)  $. $\chi_{yy}^{b}$ is the bubble
contribution, $\chi_{yy}^{l}=\chi_{yy}^{b}+\chi_{yy}^{ver}$ with $\chi
_{yy}^{ver}$ the ladder vertex correction, $\chi_{yy}^{sk}$ is the skew
scattering contribution, $\chi_{yy}=\chi_{yy}^{sk}+\chi_{yy}^{l}$. The
so-called Fermi sea term \cite{Streda1977} $\chi_{yy}^{II}$ equals zero in the
present case \cite{Xiao2017SOT}. Thus the UEE efficiency reads
\begin{align}
\alpha &  =-e\alpha_{R}D_{0}\left[  \frac{1}{\frac{e^{2}\epsilon_{F}}{\pi
\hbar^{2}}}\frac{mV_{0}}{\hbar^{2}}\frac{f_{sk}^{UEE}\left(  I_{1},I_{2}%
^{2}\right)  }{f_{L}\left(  I_{1},I_{2}^{2}\right)  }\right. \nonumber\\
&  \left.  +f_{in+AQ}\left(  I_{1},\tau iI_{2},I_{2}^{2}\right)  \rho
_{yy}\right]  . \label{scaling-UEE}%
\end{align}
The $f$'s depend on disorder via their arguments. The expressions of $f$'s are
given in Appendix C, with
\begin{align}
\text{\ }I_{1}  &  =\sum_{\eta}\left[  \sin^{2}\theta_{\eta}+\frac{1+\cos
^{2}\theta_{\eta}}{1+\left(  2\Delta_{\eta}\left(  \epsilon_{F}\right)
\frac{\tau}{\hbar}\right)  ^{2}}\right]  \frac{D_{\eta}\left(  \epsilon
_{F}\right)  }{4D_{0}},\nonumber\\
I_{2}  &  =-i\frac{\tau}{\hbar}\sum_{\eta}\frac{\epsilon_{I}}{1+\left(
2\Delta_{\eta}\left(  \epsilon_{F}\right)  \frac{\tau}{\hbar}\right)  ^{2}%
}\frac{D_{\eta}\left(  \epsilon_{F}\right)  }{D_{0}}. \label{I-1}%
\end{align}
Here $\cos\theta_{\eta}=\frac{\epsilon_{I}}{\Delta_{\eta}\left(  \epsilon
_{F}\right)  }$, $\sin\theta_{\eta}=\frac{\alpha_{R}k_{\eta}\left(
\epsilon_{F}\right)  }{\Delta_{\eta}\left(  \epsilon_{F}\right)  }$. The
results for Hall conductivities are also presented in Appendix C.

\begin{figure}[ptbh]
%Requires \usepackage{graphicx}
\includegraphics[width=0.3\textwidth]{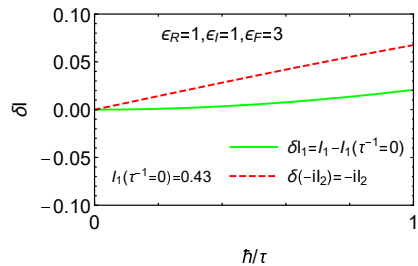} \caption{Behaviors of
$I_{1}$ and $I_{2}$ as $\hbar/\tau$ increases.}%
\label{fig1}%
\end{figure}

$I_{1,2}$ depend on the parameter $2\Delta_{\eta}\left(  \epsilon_{F}\right)
\frac{\tau}{\hbar}$ which measures the competition between the intrinsic
energy scales and the disorder-broadening of bands around the Fermi level.
When $2\Delta_{\eta}\left(  \epsilon_{F}\right)  >\hbar/\tau$, the topology of
Fermi surfaces remains unchanged and the multiband structure around the Fermi
level survives, so the Boltzmann theory is applicable. However, when
$2\Delta_{\eta}\left(  \epsilon_{F}\right)  <\hbar/\tau$, the intrinsic
multiband structure around the Fermi level collapses owing to the large
disorder-broadening. This case cannot be described by the Boltzmann theory.
More accurately, we take $\left(  \frac{\hbar}{2\Delta_{\eta}\left(
\epsilon_{F}\right)  \tau}\right)  ^{2}\leqslant0.1$, i.e., $\frac{\hbar}%
{\tau}\leqslant\frac{2\Delta_{\eta}\left(  \epsilon_{F}\right)  }{\pi}$, as
the practical criterion for the validity of Boltzmann theory. Because
$\Delta_{+}<\Delta_{-}$, $2\Delta_{+}$ is the minimal intrinsic energy scale
around the Fermi level. Thus the criterion can be refined to be $\frac{\hbar
}{\tau}\leqslant\frac{2\Delta_{+}\left(  \epsilon_{F}\right)  }{\pi}$.

\begin{figure}[ptb]
%Requires \usepackage{graphicx}
\includegraphics[width=0.34\textwidth]{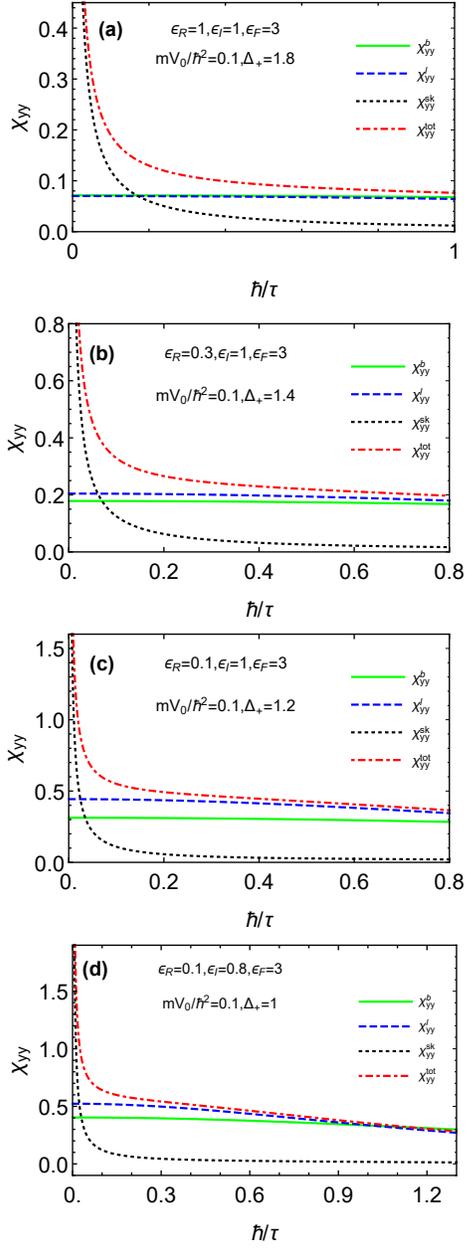} \caption{The
crossover behavior of UEE for repulsive disorder potential. $\hbar/\tau$ is
varied by tuning the impurity density $n_{im}$.}%
\label{fig2}%
\end{figure}

As shown in Fig. 1, $I_{1}$ is quite robust against increasing $\hbar/\tau$
and $I_{2}\ll I_{1}$ when $\hbar/\tau$ is smaller than $\Delta_{+}\left(
\epsilon_{F}\right)  $. When $\hbar/\tau\ll\Delta_{+}\left(  \epsilon
_{F}\right)  $ Eq. (\ref{I-1}) yields $I_{1}=C_{1}+D_{1}\left(  \frac{\hbar
}{\tau}\right)  ^{2}$ and $I_{2}=D_{2}\frac{\hbar}{\tau}$. Here $C_{1}$,
$D_{1}$ and $D_{2}$\ are disorder-independent quantities. Thus $\chi_{yy}%
^{b}\sim\tau iI_{2}$ is nearly constant when $\hbar/\tau\lesssim\frac
{2\Delta_{+}}{\pi}$, and
\[
\left(  c_{in}+c_{AQ}\right)  \propto f_{in+AQ}\left(  I_{1},\tau iI_{2}%
,I_{2}^{2}\right)  \simeq f_{in+AQ}\left(  C_{1},D_{2},0\right)  ,
\]
where $f_{in+AQ}\left(  C_{1},D_{2},0\right)  $ just corresponds to the
Boltzmann value of $\left(  c_{in}+c_{AQ}\right)  $. Thus the scaling
parameter $c_{in}+c_{AQ}$ is well-defined up to $\hbar/\tau\lesssim
\frac{2\Delta_{+}}{\pi}$. For the skew scattering,%
\[
c_{sk}\propto\frac{mV_{0}}{\hbar^{2}}\frac{f_{sk}^{UEE}\left(  I_{1},I_{2}%
^{2}\right)  }{f_{L}\left(  I_{1},I_{2}^{2}\right)  }\simeq\frac{mV_{0}}%
{\hbar^{2}}\frac{f_{sk}^{UEE}\left(  C_{1},0\right)  }{f_{L}\left(
C_{1},0\right)  }%
\]
is also expected to be insensitive to the increasing impurity density when
$\hbar/\tau\lesssim\frac{2\Delta_{+}}{\pi}$, since the corresponding Boltzmann
value is just $\frac{mV_{0}}{\hbar^{2}}\frac{f_{sk}^{UEE}\left(
C_{1},0\right)  }{f_{L}\left(  C_{1},0\right)  }$.

The definitions of the intrinsic, anomalous quantum and skew scattering
contributions are introduced in the last subsubsection in the Boltzmann
framework. Given that the Boltzmann scaling holds practically in the regime
$\hbar/\tau\lesssim\frac{2\Delta_{+}}{\pi}$, above definitions of these
contributions also remain valid in practice in this regime. Therefore, in the
case of finite but weak disorder potential, the Boltzmann scaling can be valid
even if the impurity density is not dilute in experiments. When the impurity
density increases further so that $\hbar/\tau>\frac{2\Delta_{+}}{\pi}$,
apparent $n_{im}$-dependence of $\left(  c_{in}+c_{AQ}\right)  $ is
anticipated as in Fig. 2(d), thus the Boltzmann scaling no longer work well.
In this case the conventional definitions \cite{Nagaosa2010,Sinova2015} of the
intrinsic, anomalous quantum and skew scattering contributions, which are in
fact born in the Boltzmann regime, are no longer suitable. All these points
can be read out from Fig. 4.

\subsection{General ideas based on model (\ref{model})}

\subsubsection{\textbf{Multiple intrinsic energy scales near the Fermi level}}

In complicated multiband systems there exist multiple intrinsic energy scales
around the Fermi level, e.g., $2\Delta_{+}\left(  \epsilon_{F}\right)  $ and
$2\Delta_{-}\left(  \epsilon_{F}\right)  $ in the case of both subbands
partially occupied in model (\ref{model}). The behaviors of $c_{sk}$, $c_{in}$
and $c_{AQ}$ are predominantly dictated by the smallest intrinsic energy scale
$2\Delta_{+}$. As a specific example, one can assume $2\Delta_{+}\ll
2\Delta_{-}$, then $I_{2}$ and the $\tau$-dependent part of $I_{1}$ are
dictated by $2\Delta_{+}\tau/\hbar$. This understanding accounts well for the
numerical finding in the intrinsic AHE of a multi-d-orbital tight-binding
model \cite{Kontani2007}. In Ref. \cite{Kontani2007} the minimal intrinsic
energy scale around the Fermi level is about $\Delta=0.417$ Ry (1 Ry = 13.6
eV), thus the Boltzmann scaling for the intrinsic contribution is anticipated
to be valid up to $\gamma=\frac{\hbar}{2\tau}\simeq\frac{\Delta}{2\pi}=0.066$
Ry according to our arguments. This is in exact agreement with the numerical
results presented in Ref. \cite{Kontani2007}. In the transition metal Pt it
was found that \cite{Tanaka2008} the minimal interband splitting around the
Fermi level is $\Delta=0.035$ Ry, thus the constant behavior of the intrinsic
spin Hall conductivity is anticipated to be valid up to $\gamma\simeq
\frac{\Delta}{2\pi}=5.5\times10^{-3}$ Ry, in exact agreement with the
tight-binding numerical results presented in Ref. \cite{Tanaka2008}.

The position of Fermi level dictates which intrinsic energy scales are
relevant to determining the $n_{im}$-dependence of $c_{sk}$, $c_{in}$ and
$c_{AQ}$. In model (\ref{model}), if $\epsilon_{I}\ll\epsilon_{R}$ and the
Fermi level is located within the narrow band-anticrossing region, the energy
size $2\epsilon_{I}$ of this region is the dominant intrinsic energy scale and
thus $c_{in}$ is expected to be $\tau$-independent when $\hbar/\tau
\lesssim\frac{2\epsilon_{I}}{\pi}$. This is in good agreement with the
numerical results shown in Ref. \cite{Onoda2008} ($\epsilon_{I}=0.1\ll
\epsilon_{R}\simeq1.8$), although in that paper the constant-$c_{in}$ regime
was claimed to be $\hbar/\tau\lesssim\epsilon_{F}$.

If $\epsilon_{I}>\epsilon_{R}$ and the Fermi level is located well below the
bottom of the upper subband in model (\ref{model}), as the case of Fig. 3(a),
$2\epsilon_{I}<2\Delta_{-}$ is the minimal interband splitting. The salient
feature in this case is that the intrinsic energy scale controlling
interband-coherence responses is larger than the Fermi energy (here measured
from the bottom of the lower subband). Therefore, even when $\sigma_{yy}$
(roughly proportional to $\frac{2\epsilon_{F}\tau}{\hbar}$) is not large, the
Boltzmann scaling for the AHE/SHE or UEE may still be valid. This is the case
of hole-doped MoS$_{2}$ monolayer \cite{Shan2013} with typical carrier density
$\sim10^{13}$ $cm^{-2}$, where the hole Fermi energy is smaller than the
interband splitting around the Fermi level. Assuming the Drude formula, in 3D
$\sigma_{yy}\gtrsim\frac{e^{2}}{2\pi\hbar a}\frac{\pi\epsilon_{F}}%
{\epsilon_{I}}$ in the Boltzmann regime, with $a$ the lattice constant.
Typically $\frac{e^{2}}{2\pi\hbar a}=10^{3}$ $\Omega^{-1}$ $cm^{-1}$, thus the
minimal conductivity of the Boltzmann regime is about $3\times10^{3}$
$\Omega^{-1}$ $cm^{-1}$ or smaller, if $\epsilon_{F}<\epsilon_{I}$. This
understanding provides a possible route for explaining the success of the
multivariable Boltzmann scaling for AHEs in Co$_{40}$Fe$_{40}$B$_{20}$ thin
films which were worried to be located out of the Boltzmann regime because of
the smaller conductivity $\sigma_{yy}<10^{4}$ $\Omega^{-1}cm^{-1}$
\cite{Su2014}. In the recently proposed spin-type valley Hall effect in gated
MoTe$_{2}$ described approximately by model (\ref{model}), the most pronounced
signals are obtained in the very case of Fig. 3(a) with $2\epsilon_{I}=34$
$meV$ \cite{Zhou2017}. The typical value of $\frac{\hbar}{\tau}\simeq6$ meV
\cite{Taguchi2017} is located in the Boltzmann regime $\hbar/\tau\lesssim
\frac{2\epsilon_{I}}{\pi}$, thus the Boltzmann calculation in Ref.
\cite{Zhou2017} is reliable.\begin{figure}[ptb]
%Requires \usepackage{graphicx}
\includegraphics[width=0.5\textwidth]{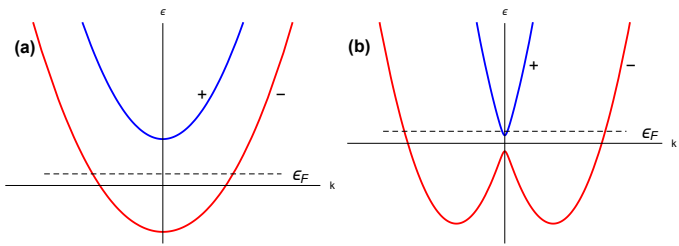} \caption{Some cases of the
Fermi-level position in model (\ref{model}) addressed in the qualitative
discussions in Sec. III. B.}%
\label{fig4}%
\end{figure}

A more subtle case occurs if other new intrinsic energy scales exist only
slightly away from the Fermi level. This case occurs also in model
(\ref{model}), as shown in Fig. 3(b): the band-anticrossing region is located
slightly away from the Fermi level. When the Fermi surface is smeared by
increasing disorder, the dominant intrinsic energy scale changes from
$2\Delta_{+}$ to $2\epsilon_{I}$. The change of the dominant intrinsic energy
scale may induce complicated behaviors of the AHE/SHE and UEE that need case
by case analysis, because the magnitude of these effects may be different for
different dominant intrinsic energy scales. In the case of Fig. 3(b), because
the intrinsic Hall current takes the largest value in the narrow
band-anticrossing region, it is expected to increase first as the
band-anticrossing region is involved when increasing disorder. After reaching
a maximum value the Hall current begins to decrease as the disorder density
increases further, because the multiband structure around the Fermi level
finally collapses. This observation accounts for the non-monotonic behavior of
the intrinsic SHE with respect to increasing $\gamma$ suggested by
tight-binding calculations in transition metal Ta \cite{Tanaka2008}. Thermal
smearing of Fermi surface has similar influences if the dominant intrinsic
energy scale is very small (%
%TCIMACRO{\TEXTsymbol{<} }%
%BeginExpansion
$<$
%EndExpansion
26 meV). Shitade et al. \cite{Shitade2012} once showed the non-monotonic
intrinsic anomalous Hall conductivity with respect to increasing temperatures
in the 2D massive Dirac model.

\subsubsection{\textbf{Multiple extrinsic energy scales}}

As we have mentioned, there are other extrinsic energy scales than
$\frac{\hbar}{\tau}$ in the case of weak disorder potential, such as
$\frac{\hbar}{\tau^{sk}}\sim\frac{\hbar}{\tau D\left\vert V_{0}\right\vert }$.
One can roughly estimate that the crossover between the skew scattering and
intrinsic-plus-anomalous-quantum (\textit{in }+\textit{ AQ}) regimes occurs at
$\frac{\hbar}{\tau^{sk}}\sim\Delta$. In the very narrow ($\epsilon_{I}%
\ll\epsilon_{R}$) resonant window of model (\ref{model}), the
skew-scattering-to-intrinsic crossover was estimated \cite{Onoda2008} to occur
at $\frac{\hbar}{\tau}\sim\frac{2m\left\vert V_{0}\right\vert }{\hbar^{2}%
}\epsilon_{I}\simeq2\pi D\left\vert V_{0}\right\vert \epsilon_{I}$, consistent
with our idea. However, out of the resonant region, we do not find a general
and rigorous theoretical criterion for the crossover. From Fig. 2, one can see
that the naively expected criterion $\frac{\hbar}{\tau}\sim2\pi D\left\vert
V_{0}\right\vert \Delta_{+}$ is only qualitatively useful, and it is likely
that other intrinsic energy scales also affects the crossover.

The $o\left(  V^{4}\right)  $ skew scattering is linked to the extrinsic
energy scale $\frac{\hbar}{\tau\left(  DV_{0}\right)  ^{2}}$ and thus is
expected to decay significantly at $\frac{\hbar}{\tau}\sim\left(
DV_{0}\right)  ^{2}\Delta$. Thus the $o\left(  V^{4}\right)  $ skew scattering
is much smaller in magnitude and decays much faster than the $o\left(
V^{3}\right)  $ one in the case of weak disorder potential. Although it is
much larger than the \textit{in }+\textit{ AQ} contribution in the limit of
dilute impurities, it is, meanwhile, overwhelmed by the $o\left(
V^{3}\right)  $ skew scattering. Thus we neglect the $o\left(  V^{4}\right)  $
skew scattering.

In this paper we only consider zero-range static impurities, for which the
transport time and quantum lifetime of electrons are not much different. This
makes the qualitative analysis of the time scale of skew scattering reliable.
If the charged impurities dominate, especially in 2D high-mobility
semiconductor heterojunctions, the ratio of the transport time and lifetime
can be very large \cite{Sarma1985}. In this case one should be cautious when
making qualitative conclusions about the skew scattering \cite{note-SKtime}.

\subsection{Qualitative phase diagram}

In the last subsection we have discussed the possible rich behaviors in the
Boltzmann regime. Now we assume the simplest case where only one dominant
intrinsic energy scale $\Delta$ is present around the Fermi level and other
intrinsic energy scales exist far away. We adopt the qualitative criterion
\cite{Onoda2008} $\frac{\hbar}{\tau}\simeq2\pi DV_{0}\frac{\Delta}{2}$ for the
crossover from the skew scattering regime to the \textit{in }+\textit{ AQ}
regime. Then we give the phase diagram in Fig. 4 for the AHE/SHE and UEE in
the case of weak disorder potential.

\begin{figure}[ptb]
%Requires \usepackage{graphicx}
\includegraphics[width=0.4\textwidth]{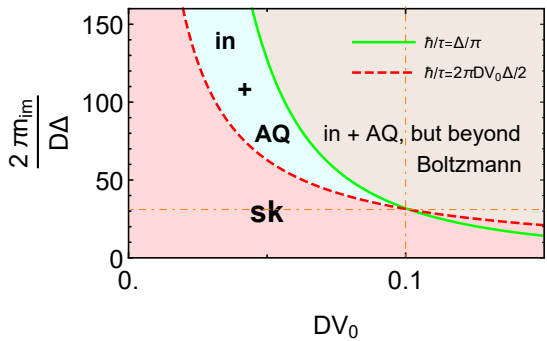} \caption{A
qualitative phase diagram for the AHE/SHE and UEE in the presence of static
impurities in the weak disorder-potential regime. The localization effect is
not included in our research. The Boltzmann scaling works well in the regime
below the green curve, and the skew scattering (sk) dominates in the regime
below the red dashed curve. In the brown regime, the
intrinsic-plus-anomalous-quantum (in + AQ) contribution still dominates, but
cannot be well described by the semiclassical Boltzmann formalism. Here we
assume $V_{0}>0$.}%
\label{fig5}%
\end{figure}

Figure 4 reveals that, in analyzing disorder effects on the AHE/SHE and UEE,
the conventional discussion based only on the dichotomy between the weak
scattering and strong scattering limits is not complete. Moreover, the usually
used term \textquotedblleft weak disorder regime\textquotedblright\ is not
clearly defined. Instead, the strength of the disorder potential and the
impurity density should be considered independently. And thus one should
distinguish the \textquotedblleft weak disorder-potential
regime\textquotedblright\ and \textquotedblleft dilute impurity
regime\textquotedblright. In the present paper we have focused on the weak
disorder-potential regime, whereas we comment on the dilute-impurity and
strong disorder-potential case in the last paragraph of this section. The
x-axis label of of Fig. 4 measures the strength of the disorder potential,
whereas the y-axis label measures the impurity density. The sk-to-\textit{in
}+\textit{ AQ} crossover is represented qualitatively by the red dashed curve,
whereas the green curve is the boundary of the Boltzmann and non-Boltzmann
regimes. We expect that this qualitative phase diagram provides a necessary
clarification of the way of thinking about disorder effects on the AHE/SHE and UEE.

There is a regime where the Boltzmann scaling holds and the \textit{in
}+\textit{ AQ} contribution dominates, thus the constant behavior of $\left(
c_{in}+c_{AQ}\right)  $ with varying $n_{im}$ is possible in experiments.
Larger $D\left\vert V_{0}\right\vert $ shrinks the range of $n_{im}$ in which
the constant behavior of $\left(  c_{in}+c_{AQ}\right)  $ may exist.

In order to address the regime $\frac{\hbar}{\tau}>\frac{\Delta}{\pi}$, one
may try to employ the equivalence $c_{in}=\chi_{yy}^{b}+\chi_{yy}^{II}$,
$c_{AQ}=\chi_{yy}^{ver}$ and $c_{sk}=\chi_{yy}^{sk}\rho_{yy}$ in the Boltzmann
regime to \textquotedblleft continue\textquotedblright\ $c_{in}$, $c_{AQ}$ and
$c_{sk}$ out of the Boltzmann regime, similar to the analytical continuation
in complex analysis. Along this route, according to Eq. (\ref{scaling-UEE})
one can view $\chi_{yy}^{b}+\chi_{yy}^{II}$, $\chi_{yy}^{ver}$ and $\chi
_{yy}^{sk}$ as the intrinsic, anomalous quantum and skew scattering
contributions, respectively, when $\frac{\hbar}{\tau}>\frac{\Delta}{\pi}$ in
the weak disorder-potential regime \cite{note-continuation}. This
\textquotedblleft continuation\textquotedblright\ has already been widely
employed in discussing the intrinsic AHE/SHE, e.g., in Refs.
\cite{Kovalev2009,Kontani2007,Kontani2007Pt,Tanaka2008}. In the weak
disorder-potential regime the continuation for the anomalous quantum and skew
scattering contributions is also feasible. As shown in Fig. 2(d), when
$\frac{\hbar}{\tau}\gtrsim\frac{2\Delta_{+}}{\pi}$ ($2\Delta_{+}=\Delta$),
even if the \textit{in }+\textit{ AQ} contribution dominates $\chi_{yy}$, one
cannot observe the well-defined Boltzmann scaling or $\tau$-independent
$\left(  c_{in}+c_{AQ}\right)  $. This situation is represented by the regime
above the green curve in Fig. 4 and most relevant in the case of very small
$\Delta$ \cite{Shitade2012,Onoda2006,Onoda2008,Kovalev2009} or very high
impurity density \cite{Tian2009}.

In the case of $DV_{0}\gtrsim0.1$ in Fig. 4, the skew scattering always
dominates over the \textit{in }+\textit{ AQ} contribution in the Boltzmann
regime. When $\frac{n_{im}}{D\Delta}$ increases into the non-Boltzmann regime
$\hbar/\tau>\frac{\Delta}{\pi}$, the \textit{in }+\textit{ AQ }contribution
gradually dominates over the skew scattering, but meanwhile the semiclassical
Boltzmann formalism already breaks down and one cannot observe $n_{im}%
$-independent constant $\left(  c_{in}+c_{AQ}\right)  $.

Before ending this section, we mention that Luttinger and Kohn also designed a
transport formalism in the dilute impurity limit without limiting the strength
of disorder potential, based on a impurity-density expansion \cite{LK1958}.
The Boltzmann equation for free electrons ($\epsilon_{k}=\frac{\hbar^{2}k^{2}%
}{2m}$) was produced, from which the nonequilibrium distribution function of
leading order $o\left(  n_{im}^{-1}\right)  $ and sub-leading order $o\left(
n_{im}^{0}\right)  $\ can be obtained \cite{LK1958}. Thus it is anticipated
that when the impurity density is low but finite, the Boltzmann formalism is
still valid if the disorder potential is not too strong. This is consistent
with the trend of our qualitative phase diagram in the larger-$DV_{0}$ part in
Fig. 4. Accordingly we speculate that the rich transport physics, such as the
crossover from the skew scattering to \textit{in} +\textit{ AQ} regime, in the
case of strong disorder-potential mainly occurs out of the Boltzmann regime.
Nevertheless, a comprehensive picture for the anomalous quantum contribution
in the Boltzmann formalism in the strong disorder-potential and dilute
impurity case is still absent. This issue calls for more future attention.

\section{Summary}

In summary, we have extended the semiclassical Boltzmann formalism for the AHE
to SHE and UEE, and confirmed this semiclassical formalism by extending the
Kohn-Luttinger density-matrix transport approach in the weak
disorder-potential regime to the linear response of spin current and spin
density. Then we investigated the regime of validity of the Boltzmann scaling
for the AHE/SHE and UEE, by performing Kubo linear response calculations in a
simple but prototypical multiband independent-carrier (electron or hole)
model. It is found that the Boltzmann scaling is practically valid provided
that the disorder-broadening of bands is quite smaller than the minimal
intrinsic energy-scale around the Fermi level. We also illustrated that the
qualitative insights acquired in the prototypical model system indeed account
for the behaviors of the AHE/SHE in various realistic systems. Moreover, we
proposed a qualitative phase diagram showing the influences of changing
independently the impurity density and the strength of disorder potential on
the AHE/SHE and UEE.

\begin{acknowledgments}
We acknowledge insightful discussions with Q. Niu. C. X. and B. X. are supported by DOE (DE-FG03-02ER45958, Division of Materials Science and Engineering), NSF (EFMA-1641101) and Welch Foundation (F-1255).
F. X. is supported by the Department of Energy, Office of Basic Energy Sciences under Contract No. DE-FG02-ER45958 and by the Welch foundation under Grant No. TBF1473.
The analysis in Sec. III is supported by the DOE grant.
\end{acknowledgments}

\appendix

\section{Justification of Eq. (\ref{semi-A-inter})}

We only address the part of the Kohn-Luttinger density-matrix approach that is
necessary for confirming our Eq. (\ref{semi-A-inter}). The original
Kohn-Luttinger approach deals with the response of electric current to the
external electric field. Here we just extend it to other observables such as
spin current and spin density. This extension only concerns the response of
off-diagonal elements of the single-particle density matrix to the external
weak dc uniform electric field.

\subsection{Preliminaries: Kohn-Luttinger single-particle formulation and
linear response}

We introduce the notation $\tilde{A}$ to mean the representation of operator
$\hat{A}$ in the second-quantized formalism. For a single-carrier operator,
i.e., $\hat{A}=\sum_{i}\hat{A}_{i}$ where $\hat{A}_{i}$ depends only on the
dynamical variables of the $i$-th carrier, one has $\tilde{A}=\sum
_{nn^{\prime}}A_{nn^{\prime}}a_{n}^{\dag}a_{n^{\prime}}$ where $A_{nn^{\prime
}}$ are the matrix elements in the $n$ representation of single-carrier space,
$a_{n}^{\dag}$ ($a_{n}$) is the creation (annihilation) operator on the
single-carrier eigenstate $\left\vert n\right\rangle $. The expectation value
of $\hat{A}$ is given by $\left\langle A\right\rangle =Tr\left(  \tilde{\rho
}_{T}\tilde{A}\right)  $, where $Tr$ denotes the trace operation in the
occupation-number space, and the many-particle density matrix $\tilde{\rho
}_{T}$ in the occupation-number representation is governed by the quantum
Liouville equation $i\hbar\frac{\partial}{\partial t}\tilde{\rho}_{T}=\left[
\tilde{H}_{T},\tilde{\rho}_{T}\right]  $.

The expectation value of a single-carrier operator $\hat{A}$ can be expressed
in terms of $\hat{A}$ and a single-carrier operator $\hat{\rho}_{T}$:
\begin{align}
\left\langle A\right\rangle  &  =\sum_{nn^{\prime}}A_{nn^{\prime}}\left(
\hat{\rho}_{T}\right)  _{n^{\prime}n}=tr\left[  \hat{A}\hat{\rho}_{T}\right]
,\text{ }\nonumber\\
\left(  \hat{\rho}_{T}\right)  _{n^{\prime}n}  &  \equiv Tr\left(  \tilde
{\rho}_{T}a_{n}^{\dag}a_{n^{\prime}}\right)  .
\end{align}
Here $tr$ denotes the trace operation in the single-carrier Hilbert space.
Kohn-Luttinger noticed that \cite{KL1957}, when the total Hamiltonian is also
a single-carrier operator $\tilde{H}_{T}=\sum_{mm^{\prime}}\left(  \hat{H}%
_{T}\right)  _{mm^{\prime}}a_{m}^{\dag}a_{m^{\prime}}$, the equation of motion
for $\left(  \hat{\rho}_{T}\right)  _{n^{\prime}n}$ reads (Schroedinger
picture) $i\hbar\frac{\partial}{\partial t}\left(  \hat{\rho}_{T}\right)
_{n^{\prime}n}=\left[  \hat{H}_{T},\hat{\rho}_{T}\right]  _{n^{\prime}n}$. The
$n$ representation in the single-carrier Hilbert space is arbitrary thus
\begin{equation}
i\hbar\frac{\partial}{\partial t}\hat{\rho}_{T}=\left[  \hat{H}_{T},\hat{\rho
}_{T}\right]  \label{single-particle QLE}%
\end{equation}
with the operators acting on the single-carrier space. $\hat{\rho}_{T}$
satisfies $\left(  \hat{\rho}_{T}\right)  _{nn}=\left\langle N_{n}%
\right\rangle \geq0$ and $tr\hat{\rho}_{T}=N_{c}$ with $N_{n}=a_{n}^{\dag
}a_{n}$ and $\tilde{N}=\sum_{n}N_{n}$. Although normalized to the carrier
number $N_{c}$ instead of 1, $\hat{\rho}_{T}$ is often referred to as the
single-particle density matrix, the diagonal elements of which represent the
average occupation numbers of single-particle eigenstates rather than
occupation probability. This character implies that $\hat{\rho}_{T}$ can be
regarded as a quantum-statistical generalization of the single-particle
density function described by the classical Boltzmann equation, and thus the
equation of motion for $\hat{\rho}_{T}$ may reduce to a Boltzmann-type
transport equation for diagonal elements of $\hat{\rho}_{T}$. This idea
motivates one to split the quantum Liouville equation in the band-eigenstate
representation into diagonal and off-diagonal parts in the following.

The single-carrier Hamiltonian reads $\hat{H}_{T}=\hat{H}_{0}+\hat{H}^{\prime
}+\hat{H}_{F}$, where $\hat{H}_{0}$ is the single-particle free Hamiltonian,
$\hat{H}^{\prime}=\lambda\hat{V}$ with $\lambda$ a dimensionless parameter and
$\hat{V}$ the potential produced by randomly distributed static impurities,
and the field term $\hat{H}_{F}=\hat{H}_{1}e^{st}$ with $\hat{H}%
_{1}=-e\mathbf{E\cdot r}$ arises from the electric field adiabatically
switched-on from the remote past $t=-\infty$. The infinitesimal positive $s$
in $\hat{H}_{F}$ can be taken to be the same as the $s$ which appears as a
regularization factor in the T-matrix theory of the semiclassical Boltzmann
formalism (see the main text). This is because the physical situation is
obtained by taking the limit $s\rightarrow0^{+}$. We remind that a similar
note on the infinitesimal positive $s$ has appeared in the derivation of
Kubo-Streda linear response formula with respect to the uniform static
electric field \cite{Streda2010}.

The Kohn-Luttinger theory starts from Eq. (\ref{single-particle QLE}) in the
linear response regime where $\hat{\rho}_{T}=\hat{\rho}+\hat{\rho}_{F}$. Here
$\hat{\rho}$ is the equilibrium value of the single-particle density matrix,
$\hat{\rho}_{F}$ is linear in the electric field and satisfies $\hat{\rho}%
_{F}\left(  t\rightarrow-\infty\right)  =0$. Kohn and Luttinger proceeded by
employing the ansatz $\hat{\rho}_{F}=\hat{f}e^{st}$, where $\hat{f}=\hat{\rho
}_{F}\left(  t=0\right)  $ is independent of time. Then Eq.
(\ref{single-particle QLE}) reduces to
\begin{equation}
d_{ll^{\prime}}^{-}f_{ll^{\prime}}=\sum_{l^{\prime\prime}}\left(
f_{ll^{\prime\prime}}H_{l^{\prime\prime}l^{\prime}}-H_{ll^{\prime\prime}%
}f_{l^{\prime\prime}l^{\prime}}\right)  +C_{ll^{\prime}} \label{KL}%
\end{equation}
in the band-eigenstate representation of $\hat{H}_{0}$. Here $C_{ll^{\prime}%
}\equiv\left[  \hat{\rho},\hat{H}_{1}\right]  _{ll^{\prime}}$ reads
$C_{ll^{\prime}}=ie\mathbf{E}\cdot\left[  \left(  \partial_{\mathbf{k}%
}+\partial_{\mathbf{k}^{\prime}}\right)  \rho_{ll^{\prime}}+\left[
\mathbf{J},\rho\right]  _{ll^{\prime}}\right]  $ for $l\neq l^{\prime}$, and
$C_{l}=ie\mathbf{E}\cdot\left[  \partial_{\mathbf{k}}\rho_{ll}+\left[
\mathbf{J},\rho\right]  _{ll}\right]  $, where $\left[  \mathbf{J}%
,\rho\right]  _{ll^{\prime}}\equiv\sum_{l^{\prime\prime}}\left(
\mathbf{J}_{ll^{\prime\prime}}\rho_{l^{\prime\prime}l^{\prime}}-\rho
_{ll^{\prime\prime}}\mathbf{J}_{l^{\prime\prime}l^{\prime}}\right)  $. Here
$\mathbf{r}_{ll^{\prime}}=i\frac{\partial}{\partial\mathbf{k}}\delta
_{ll^{\prime}}+i\mathbf{J}_{ll^{\prime}}$\ and $\mathbf{J}_{ll^{\prime}}%
\equiv\delta_{\mathbf{kk}^{\prime}}\langle u_{l}|\partial_{\mathbf{k}%
}|u_{l^{\prime}}\rangle$ are used.

The linear response of an observable $A$ is thus%
\begin{equation}
\delta A=tr\left\langle \hat{f}\hat{A}\right\rangle =\sum_{l}\left\langle
f_{l}\right\rangle A_{ll}+\sum_{ll^{\prime}}^{\prime}\left\langle
f_{ll^{\prime}}\right\rangle A_{l^{\prime}l},
\end{equation}
Hereafter the notation $\sum^{\prime}$ means that all the index equalities
should be avoided in the summation.

Equation (\ref{KL}) can be split into
\begin{equation}
d_{ll^{\prime}}^{-}f_{ll^{\prime}}=\sum_{l^{\prime\prime}}^{\prime}\left(
f_{ll^{\prime\prime}}H_{l^{\prime\prime}l^{\prime}}^{\prime}-H_{ll^{\prime
\prime}}^{\prime}f_{l^{\prime\prime}l^{\prime}}\right)  +\left(
f_{l}-f_{l^{\prime}}\right)  H_{ll^{\prime}}^{\prime}+C_{ll^{\prime}}
\label{KL-offdiagonal}%
\end{equation}
for $l\neq l^{\prime}$, and
\begin{equation}
-i\hbar sf_{l}=\sum_{l^{\prime}\neq l}\left(  f_{ll^{\prime}}H_{l^{\prime}%
l}^{\prime}-H_{ll^{\prime}}^{\prime}f_{l^{\prime}l}\right)  +C_{l}.
\label{KL-diagonal}%
\end{equation}
Here $H_{ll}^{\prime}$, which is the first-order energy correction in the bare
quantum mechanical perturbation theory, has been absorbed into $H_{0}$, thus
$H_{ll}^{\prime}=0$ hereafter. In the case of weak disorder-potential, an
iterative analysis of Eqs. (\ref{KL-offdiagonal}) and (\ref{KL-diagonal}) in
terms of the parameter $\lambda$ is possible. Assuming $sf_{l}\rightarrow0$
when $s\rightarrow0^{+}$ is equivalent to assuming that $f_{l}$\ starts from
the order of $\lambda^{-2}$ \cite{KL1957,Moore1967,Nagaosa2010}.\ Then an
order-by-order analysis with respect to the disorder potential follows:%
\begin{align*}
f_{l}  &  =f_{l}^{\left(  -2\right)  }+f_{l}^{\left(  -1\right)  }%
+f_{l}^{\left(  0\right)  }+...,\\
f_{ll^{\prime}}  &  =f_{ll^{\prime}}^{\left(  -1\right)  }+f_{ll^{\prime}%
}^{\left(  0\right)  }+f_{ll^{\prime}}^{\left(  1\right)  }...\text{\ }\left(
l\neq l^{\prime}\right)  ,\\
C_{ll^{\prime}}  &  =C_{ll^{\prime}}^{\left(  0\right)  }+C_{ll^{\prime}%
}^{\left(  1\right)  }+C_{ll^{\prime}}^{\left(  2\right)  }+...
\end{align*}
The superscript means the order of $\lambda$. The iterative solutions are not
repeated here. Then one gets the expressions for the off-diagonal elements in
terms of the diagonal ones \cite{Luttinger1958}, e.g.,%
\[
f_{ll^{\prime}}^{\left(  -1\right)  }=\frac{f_{l}^{\left(  -2\right)
}-f_{l^{\prime}}^{\left(  -2\right)  }}{d_{ll^{\prime}}^{-}}H_{ll^{\prime}%
}^{\prime},
\]%
\begin{align*}
f_{ll^{\prime}}^{\left(  0\right)  }  &  =\sum_{l^{\prime\prime}\neq
l,l^{\prime}}\frac{H_{ll^{\prime\prime}}^{\prime}H_{l^{\prime\prime}l^{\prime
}}^{\prime}}{d_{ll^{\prime}}^{-}}\left[  \frac{f_{l}^{\left(  -2\right)
}-f_{l^{\prime\prime}}^{\left(  -2\right)  }}{d_{ll^{\prime\prime}}^{-}}%
-\frac{f_{l^{\prime\prime}}^{\left(  -2\right)  }-f_{l^{\prime}}^{\left(
-2\right)  }}{d_{l^{\prime\prime}l^{\prime}}^{-}}\right] \\
&  +\frac{f_{l}^{\left(  -1\right)  }-f_{l^{\prime}}^{\left(  -1\right)  }%
}{d_{ll^{\prime}}^{-}}H_{ll^{\prime}}^{\prime}+\frac{C_{ll^{\prime}}^{\left(
0\right)  }}{d_{ll^{\prime}}^{-}},
\end{align*}
and a transport equation which only concerns the diagonal elements. These
equations are microscopic equations, and the required macroscopic equations
are obtained from them by disorder-averaging. In so doing, Kohn and Luttinger
assumed that $f_{l}$ does not contain any physically important, rapidly
varying exponential factors, thus in the thermodynamic limit $\left\langle
f_{l}^{\left(  -2\right)  }H_{ll^{\prime}}^{\prime}\right\rangle =\left\langle
f_{l}^{\left(  -2\right)  }\right\rangle \left\langle H_{ll^{\prime}}^{\prime
}\right\rangle $, $\left\langle H_{ll^{\prime\prime}}^{\prime}H_{l^{\prime
\prime}l^{\prime}}^{\prime}f_{l}^{\left(  -2\right)  }\right\rangle
=\left\langle H_{ll^{\prime\prime}}^{\prime}H_{l^{\prime\prime}l^{\prime}%
}^{\prime}\right\rangle \left\langle f_{l}^{\left(  -2\right)  }\right\rangle
$. Therefore, one can see that, only when this assumption is true, the
semiclassical distribution function and thus the Boltzmann formalism can be
defined. The validity of this vital assumption has been confirmed by
subsequent researches \cite{Moore1967}, but beyond the scope of our study.

In the case of weak disorder-potential, the off-diagonal response only
concerns the lowest nonzero order of $\left\langle f_{ll^{\prime}%
}\right\rangle $ (see below), while the analysis of diagonal response has to
go to higher orders in the perturbation expansion of $f_{l}$. In these
higher-order contributions some trivial renormalization effects appear
\cite{KL1957,Luttinger1958,Moore1967}, only giving rise to negligible
higher-order contributions in the weak disorder-potential limit to AHE/SHE and
UEE \cite{Sinitsyn2006,Sinitsyn2008}. The qualitatively and quantitatively
important part of the diagonal response of density matrix in the weak
disorder-potential regime is just the generalized semiclassical Boltzmann equation.

\subsection{Off-diagonal response}

After disorder average, assuming $\left\langle H_{ll^{\prime}}^{\prime
}\right\rangle =0$ one has $\left\langle f_{ll^{\prime}}^{\left(  -1\right)
}\right\rangle =0$ and%
\begin{gather}
\sum_{ll^{\prime}}^{\prime}\left\langle f_{ll^{\prime}}\right\rangle
A_{l^{\prime}l}=\sum_{ll^{\prime}}^{\prime}\left\langle f_{ll^{\prime}%
}^{\left(  0\right)  }\right\rangle A_{l^{\prime}l}=\sum_{ll^{\prime}}%
^{\prime}C_{ll^{\prime}}^{\left(  0\right)  }\frac{A_{l^{\prime}l}%
}{d_{ll^{\prime}}^{-}}+\nonumber\\
\sum_{ll^{\prime}l^{\prime\prime}}^{\prime}\left\langle \frac{f_{l}^{\left(
-2\right)  }-f_{l^{\prime\prime}}^{\left(  -2\right)  }}{d_{ll^{\prime\prime}%
}^{-}}-\frac{f_{l^{\prime\prime}}^{\left(  -2\right)  }-f_{l^{\prime}%
}^{\left(  -2\right)  }}{d_{l^{\prime\prime}l^{\prime}}^{-}}\right\rangle
\frac{\left\langle H_{ll^{\prime\prime}}^{\prime}H_{l^{\prime\prime}l^{\prime
}}^{\prime}\right\rangle A_{l^{\prime}l}}{d_{ll^{\prime}}^{-}}.
\end{gather}
Due to $C_{ll^{\prime}}^{\left(  0\right)  }=ie\mathbf{E\cdot J}_{ll^{\prime}%
}\left(  \rho_{l^{\prime}}-\rho_{l}\right)  $ and $\mathbf{v}_{ll^{\prime}%
}\delta_{\mathbf{kk}^{\prime}}=-\frac{1}{\hbar}d_{ll^{\prime}}\mathbf{J}%
_{ll^{\prime}}$ for $l\neq l^{\prime}$, we have%
\begin{align}
\sum_{ll^{\prime}}^{\prime}C_{ll^{\prime}}^{\left(  0\right)  }\frac
{A_{l^{\prime}l}}{d_{ll^{\prime}}^{-}}  &  =2e\sum_{ll^{\prime}}^{\prime}%
\rho_{l}\mathrm{Im}\frac{\mathbf{E}\cdot\mathbf{J}_{ll^{\prime}}A_{l^{\prime
}l}}{d_{ll^{\prime}}}\nonumber\\
&  =-2\hbar e\sum_{ll^{\prime}}^{\prime}\rho_{l}\delta_{\mathbf{kk}^{\prime}%
}\frac{\mathrm{Im}\mathbf{E\cdot}\langle u_{l}|\mathbf{v}|u_{l^{\prime}%
}\rangle A_{l^{\prime}l}}{d_{ll^{\prime}}^{2}}\nonumber\\
&  \equiv\sum_{l}f_{l}^{0}\delta^{in}A_{l},
\end{align}
where $\rho_{l}=f_{l}^{0}$.

Besides, by interchanging the indices $l$, $l^{\prime}$ and $l^{\prime\prime}$
here and there and some simple algebra, we find%
\begin{gather}
\sum_{ll^{\prime}l^{\prime\prime}}^{\prime}\left\langle H_{ll^{\prime\prime}%
}^{\prime}H_{l^{\prime\prime}l^{\prime}}^{\prime}\right\rangle \left\langle
\frac{f_{l}^{\left(  -2\right)  }-f_{l^{\prime\prime}}^{\left(  -2\right)  }%
}{d_{ll^{\prime\prime}}^{-}}-\frac{f_{l^{\prime\prime}}^{\left(  -2\right)
}-f_{l^{\prime}}^{\left(  -2\right)  }}{d_{l^{\prime\prime}l^{\prime}}^{-}%
}\right\rangle \frac{A_{l^{\prime}l}}{d_{ll^{\prime}}^{-}}\nonumber\\
=\sum_{ll^{\prime}l^{\prime\prime}}^{\prime}\left\langle f_{l}^{\left(
-2\right)  }\right\rangle \left[  \frac{\left\langle H_{l^{\prime}%
l^{\prime\prime}}^{\prime}H_{l^{\prime\prime}l}^{\prime}\right\rangle
A_{ll^{\prime}}}{d_{ll^{\prime}}^{+}d_{ll^{\prime\prime}}^{+}}+c.c.\right]
\nonumber\\
+\sum_{ll^{\prime}l^{\prime\prime}}^{\prime}\left\langle f_{l}^{\left(
-2\right)  }\right\rangle \left\langle H_{l^{\prime\prime}l}^{\prime
}H_{ll^{\prime}}^{\prime}\right\rangle A_{l^{\prime}l^{\prime\prime}}\left(
\frac{1}{d_{ll^{\prime\prime}}^{+}}-\frac{1}{d_{ll^{\prime}}^{-}}\right)
\frac{1}{d_{l^{\prime\prime}l^{\prime}}^{-}}\nonumber\\
=\sum_{ll^{\prime}l^{\prime\prime}}^{\prime}\left\langle f_{l}^{\left(
-2\right)  }\right\rangle \left[  2\operatorname{Re}\frac{\left\langle
H_{l^{\prime}l^{\prime\prime}}^{\prime}H_{l^{\prime\prime}l}^{\prime
}\right\rangle A_{ll^{\prime}}}{d_{ll^{\prime}}^{+}d_{ll^{\prime\prime}}^{+}%
}+\frac{\left\langle H_{l^{\prime\prime}l}^{\prime}H_{ll^{\prime}}^{\prime
}\right\rangle A_{l^{\prime}l^{\prime\prime}}}{d_{ll^{\prime\prime}}%
^{+}d_{ll^{\prime}}^{-}}\right] \nonumber\\
=\sum_{l}\left\langle f_{l}^{\left(  -2\right)  }\right\rangle \delta
^{sj}A_{l},
\end{gather}
where $\delta^{sj}A_{l}$ coincides with Eq. (\ref{ex}).

Summarizing the contents of this subsection, we proved in the weak
disorder-potential regime and linear response regime%
\begin{equation}
\sum_{ll^{\prime}}^{\prime}\left\langle f_{ll^{\prime}}\right\rangle
A_{l^{\prime}l}=\sum_{l}f_{l}^{0}\delta^{in}A_{l}+\sum_{l}\left\langle
f_{l}^{\left(  -2\right)  }\right\rangle \delta^{sj}A_{l}.
\end{equation}
Here $\left\langle f_{l}^{\left(  -2\right)  }\right\rangle $ is just the
conventional nonequilibrium distribution function obtained in the lowest Born
approximation. This equation confirms what was obtained
semi-phenomenologically previously in the Boltzmann formalism
\cite{Xiao2017SOT-SBE}.

\section{Scattering off pairs of impurities}

In the modern semiclassical Boltzmann theory developed in studying the AHE,
the anti-symmetric part $\omega_{ll^{\prime}}^{4a}\equiv\frac{1}{2}\left(
\omega_{ll^{\prime}}^{\left(  4\right)  }-\omega_{l^{\prime}l}^{\left(
4\right)  }\right)  $ of the fourth-order scattering rate $\omega_{ll^{\prime
}}^{\left(  4\right)  }$ was calculated only within the noncrossing
approximation, giving rise to the intrinsic-skew-scattering-induced anomalous
quantum contribution \cite{Sinitsyn2008}. Here we show that the crossing part
also contributes to the anomalous quantum contribution. Both the noncrossing
and crossing contributions arise from scattering off pairs of impurity centers
\cite{Ado}.

Starting from \cite{Xiao2017AHE}
\begin{align*}
\omega_{ll^{\prime}}^{4a}  &  =-\frac{2\pi}{\hbar}\delta\left(  d_{ll^{\prime
}}\right)  \sum_{l^{\prime\prime},l^{\prime\prime\prime}}^{\prime}\left[
\mathrm{Im}\left\langle V_{ll^{\prime\prime\prime}}V_{l^{\prime\prime\prime
}l^{\prime}}V_{l^{\prime}l^{\prime\prime}}V_{l^{\prime\prime}l}\right\rangle
\mathrm{Im}\frac{1}{d_{ll^{\prime\prime}}^{-}d_{ll^{\prime\prime\prime}}^{+}%
}\right. \\
&  \left.  +\mathrm{Im}\left\langle V_{ll^{\prime}}V_{l^{\prime}%
l^{\prime\prime}}V_{l^{\prime\prime}l^{\prime\prime\prime}}V_{l^{\prime
\prime\prime}l}\right\rangle \mathrm{Im}\frac{1}{d_{ll^{\prime\prime}}%
^{-}d_{ll^{\prime\prime\prime}}^{-}}\right. \\
&  \left.  +\mathrm{Im}\left\langle V_{ll^{\prime\prime\prime}}V_{l^{\prime
\prime\prime}l^{\prime\prime}}V_{l^{\prime\prime}l^{\prime}}V_{l^{\prime}%
l}\right\rangle \mathrm{Im}\frac{1}{d_{ll^{\prime\prime}}^{+}d_{ll^{\prime
\prime\prime}}^{+}}\right]  ,
\end{align*}
we get
\begin{align*}
\omega_{ll^{\prime}}^{4a}  &  =-\frac{\left(  2\pi\right)  ^{2}}{\hbar}%
\delta\left(  d_{ll^{\prime}}\right)  \sum_{l^{\prime\prime},l^{\prime
\prime\prime}}^{\prime}\frac{\delta\left(  d_{ll^{\prime\prime}}\right)
}{d_{ll^{\prime\prime\prime}}}\left[  \mathrm{Im}\left\langle V_{ll^{\prime
\prime\prime}}V_{l^{\prime\prime\prime}l^{\prime}}V_{l^{\prime}l^{\prime
\prime}}V_{l^{\prime\prime}l}\right\rangle \right. \\
&  \left.  +\mathrm{Im}\left\langle V_{ll^{\prime}}V_{l^{\prime}%
l^{\prime\prime}}V_{l^{\prime\prime}l^{\prime\prime\prime}}V_{l^{\prime
\prime\prime}l}\right\rangle +\mathrm{Im}\left\langle V_{ll^{\prime}%
}V_{l^{\prime}l^{\prime\prime\prime}}V_{l^{\prime\prime\prime}l^{\prime\prime
}}V_{l^{\prime\prime}l}\right\rangle \right]  ,
\end{align*}
where the $l,l^{\prime}$ and $l^{\prime\prime}$ states are on-shell, whereas
the $l^{\prime\prime\prime}$ state may be off--shell. When taking the average
over disorder configurations, there exist some different possibilities: the
non-Gaussian contribution from $o\left(  V^{4}\right)  $ disorder correlation
\cite{Kovalev2008,Kovalev2009,Xiao2017AHE}, the Gaussian non-crossing
\cite{Sinitsyn2007} and crossing \cite{Ado} contributions. For the noncrossing
Gaussian contribution, one can find that an interband off-shell scattering is
contained in each term. While for crossing Gaussian contribution, the
interband off-shell scattering may be present or not. To be more specific, we
present the expressions for $\omega_{ll^{\prime}}^{4a}$ in the smooth scalar
disorder-potential limit. In this limit, the two momenta linked by the
disorder potential are close to each other thus
\begin{align*}
V_{ll^{\prime}}  &  =V_{\mathbf{kk}^{\prime}}\left[  \delta_{\eta\eta^{\prime
}}+\left(  k_{\mu}^{\prime}-k_{\mu}\right)  J_{\mu}^{\eta\eta^{\prime}}\left(
\mathbf{k}\right)  \right. \\
&  \left.  +\frac{1}{2}\left(  k_{\mu}^{\prime}-k_{\mu}\right)  \left(
k_{\nu}^{\prime}-k_{\nu}\right)  J_{\mu\nu}^{\eta\eta^{\prime}}\left(
\mathbf{k}\right)  +...\right]  ,
\end{align*}
where $J_{\mu}^{\eta\eta^{\prime}}\left(  \mathbf{k}\right)  =\langle
u_{\eta\mathbf{k}}|\partial_{k_{\mu}}|u_{\eta^{\prime}\mathbf{k}}\rangle$ and
$J_{\mu\nu}^{\eta\eta^{\prime}}\left(  \mathbf{k}\right)  =\langle
u_{\eta\mathbf{k}}|\partial_{k_{\nu}}\partial_{k_{\mu}}|u_{\eta^{\prime
}\mathbf{k}}\rangle$. The Einstein summation convention is used hereafter for
the indices $\mu$, $\nu$. For real elastic process there must be $\eta
=\eta^{\prime}$ in the smooth scalar disorder-potential limit in
non-degenerate multiband system. When taking the disorder average, we only
consider Gaussian disorder. The noncrossing part contributes%
\begin{align*}
\omega_{ll^{\prime}}^{4a-nc}  &  =\frac{\left(  2\pi n_{im}V_{0}^{2}\right)
^{2}}{2\hbar}\left(  \mathbf{k\times k}^{\prime}\right)  _{\mu\nu}\delta
_{\eta^{\prime}\eta}\delta\left(  d_{ll^{\prime}}\right) \\
&  \times\sum_{l^{\prime\prime},l^{\prime\prime\prime}}^{\prime}\frac
{\delta_{\eta^{\prime\prime}\eta}\delta\left(  d_{ll^{\prime\prime}}\right)
}{d_{ll^{\prime\prime\prime}}}\left(  \delta_{\mathbf{k}^{\prime\prime
}\mathbf{k}^{\prime\prime\prime}}+\delta_{\mathbf{k}^{\prime}\mathbf{k}%
^{\prime\prime\prime}}+\delta_{\mathbf{kk}^{\prime\prime\prime}}\right) \\
&  \times\mathrm{Im}\left[  J_{\mu}^{\eta\eta^{\prime\prime\prime}}\left(
\mathbf{k}\right)  J_{\nu}^{\eta^{\prime\prime\prime}\eta}\left(
\mathbf{k}\right)  \right]  .
\end{align*}
Thus an interband off-shell scattering ($\eta^{\prime\prime\prime}\neq\eta$)
is unavoidable in each term of the noncrossing
intrinsic-skew-scattering-induced anomalous quantum contribution
\cite{Sinitsyn2007,Kovalev2010}. For the crossing
coherent-skew-scattering-induced anomalous quantum contribution \cite{Ado}, we
get
\begin{align*}
\omega_{ll^{\prime}}^{4a-c}  &  =-\frac{\left(  2\pi n_{im}V_{0}^{2}\right)
^{2}}{2\hbar}\delta_{\eta^{\prime}\eta}\delta\left(  d_{ll^{\prime}}\right)
\sum_{l^{\prime\prime},l^{\prime\prime\prime}}^{\prime}\frac{\delta
_{\eta^{\prime\prime}\eta}\delta\left(  d_{ll^{\prime\prime}}\right)
}{d_{ll^{\prime\prime\prime}}}\\
&  \times\left(  \mathbf{k\times k}^{\prime}+\mathbf{k}^{\prime}%
\times\mathbf{k}^{\prime\prime}+\mathbf{k}^{\prime\prime}\times\mathbf{k}%
\right)  _{\mu\nu}\\
&  \times\left(  \delta_{\mathbf{k+\mathbf{k}}^{\prime}\mathbf{\mathbf{=k}%
}^{\prime\prime}\mathbf{+k}^{\prime\prime\prime}}+\delta_{\mathbf{k+\mathbf{k}%
}^{\prime\prime}\mathbf{\mathbf{=k}}^{\prime}\mathbf{+k}^{\prime\prime\prime}%
}+\delta_{\mathbf{k+\mathbf{k}}^{\prime\prime\prime}\mathbf{\mathbf{=k}%
}^{\prime\prime}\mathbf{+\mathbf{k}}^{\prime}}\right) \\
&  \times\left\{  \mathrm{Im}\left[  J_{\mu}^{\eta\eta^{\prime\prime\prime}%
}\left(  \mathbf{k}\right)  J_{\nu}^{\eta^{\prime\prime\prime}\eta}\left(
\mathbf{k}\right)  \right]  -\delta_{\eta^{\prime\prime\prime}\eta}\Omega
_{\mu\nu}\left(  \mathbf{k}\right)  \right\}  ,
\end{align*}
which contains both intraband ($\eta^{\prime\prime\prime}=\eta$) and interband
($\eta^{\prime\prime\prime}\neq\eta$) terms. $\Omega_{\mu\nu}\left(
\mathbf{k}\right)  $ is the momentum-space Berry curvature. In fact, this
expression was already obtained by Luttinger sixty years ago
\cite{Luttinger1958}.

\section{Calculation details}

\subsection{Boltzmann calculation}

For identical pointlike scalar impurities in model (\ref{model}) the
conventional skew scattering from $o\left(  V^{3}\right)  $ non-Gaussian
disorder vanishes due to $\omega_{ll^{\prime}}^{3a}=0$ \cite{Xiao2017AHE},
then the skew scattering induced by the $o\left(  V^{4}\right)  $ non-Gaussian
correlation $\left\langle V^{4}\right\rangle _{c}=n_{im}V_{0}^{4}$ plays an
important role in the dilute limit \cite{Kovalev2008,Kovalev2009,Xiao2017AHE}.
The nonequilibrium distribution function responsible for this $o\left(
V^{4}\right)  $ skew scattering reads\ \cite{Xiao2017AHE} $g_{\eta}%
^{sk}\left(  \epsilon\right)  =\left(  -\partial_{\epsilon}f^{0}\right)
\left(  \mathbf{\hat{z}}\times e\mathbf{E}\right)  \cdot\mathbf{v}_{\eta}%
^{0}\tau_{\eta}^{sk}\left(  \epsilon\right)  $ with
\[
\tau_{\eta}^{sk}\left(  \epsilon\right)  =-\tau\left(  \frac{mV_{0}}{\hbar
^{2}}\right)  ^{2}\frac{\eta\epsilon_{I}\epsilon_{R}\left(  \epsilon_{I}%
^{2}+2\epsilon_{R}\epsilon\right)  }{2\pi\bar{\Delta}^{2}\left(
\epsilon\right)  \left(  \epsilon_{I}^{2}+\epsilon_{R}\epsilon\right)  }%
\ln\frac{k_{-}\left(  \epsilon\right)  }{k_{+}\left(  \epsilon\right)  },
\]
then the corresponding nonequilibrium spin-polarization $\delta^{sk}%
\mathbf{S=}\sum_{l}g_{l}^{sk}\mathbf{S}_{l}^{0}$ takes the form
\[
\delta^{sk}\mathbf{S}=-e\alpha_{R}D_{0}\frac{\hbar D_{0}}{n_{im}}%
\frac{\epsilon_{I}\epsilon_{R}\epsilon_{F}\ln\left(  k_{-}/k_{+}\right)
}{\bar{\Delta}\left(  \epsilon_{I}^{2}+\epsilon_{R}\epsilon_{F}\right)  }%
E_{y}\mathbf{\hat{y}.}%
\]
Here $\tau\left(  \frac{mV_{0}}{\hbar^{2}}\right)  ^{2}=2\pi\hbar D_{0}%
/n_{im}$ is used. From this particular model case, one can extract a generic
information that the skew scattering contribution due to the $o\left(
V^{4}\right)  $ non-Gaussian disorder correlation is characterized by a
relaxation time of scale $\tau\left(  \frac{mV_{0}}{\hbar^{2}}\right)  ^{2}$.

However, in usual case the $o\left(  V^{3}\right)  $ skew scattering is
nonzero and dominates the skew scattering contribution. The peculiar property
of model (\ref{model}) with scalar short-range disorder in the case of both
subbands partially occupied thus makes it inconvenient to extract general insights.

As for the disorder model chosen in the main text, we obtain ($l=\left(
\eta,\mathbf{k}\right)  =\left(  \eta,\epsilon,\phi\right)  $)%
\[
\omega_{ll^{\prime}}^{a}=-\frac{1}{\tau}\frac{mV_{0}}{\hbar^{2}}\frac{\eta
\eta^{\prime}\sin\left(  \phi^{\prime}-\phi\right)  }{2D_{0}}\frac{\alpha
_{R}^{2}k_{\eta}\left(  \epsilon\right)  k_{\eta^{\prime}}\left(
\epsilon\right)  }{\Delta_{\eta}\left(  \epsilon\right)  \Delta_{\eta^{\prime
}}\left(  \epsilon\right)  }\delta\left(  d_{ll^{\prime}}\right)
\]
for the $o\left(  V^{3}\right)  $ skew scattering contribution. Substituting
$\omega_{ll^{\prime}}^{a}$ into the Boltzmann equation and following the
general recipe given in Ref. \cite{Xiao2017AHE}, we get
\[
\tau_{\eta}^{sk}\left(  \epsilon\right)  =-\tau\frac{mV_{0}}{\hbar^{2}}%
\frac{\eta\epsilon_{R}\left(  \epsilon_{I}^{2}+\epsilon_{R}\epsilon\right)
\left(  \epsilon_{I}^{2}+2\epsilon_{R}\epsilon\right)  }{\bar{\Delta}\left(
\epsilon\right)  \left(  \epsilon_{I}^{2}+3\epsilon_{R}\epsilon\right)  ^{2}%
}.
\]
Thus $\delta^{sk}\mathbf{S=}\sum_{l}g_{l}^{sk}\mathbf{S}_{l}^{0}$ is given by
\[
\delta^{sk}\mathbf{S}=-e\alpha_{R}D_{0}\tau\frac{mV_{0}}{\hbar^{2}}%
\frac{\epsilon_{R}\epsilon_{F}\left(  \epsilon_{I}^{2}+\epsilon_{R}%
\epsilon_{F}\right)  }{\left(  \epsilon_{I}^{2}+3\epsilon_{R}\epsilon
_{F}\right)  ^{2}}E_{y}\mathbf{\hat{y}.}%
\]

\subsection{Kubo calculation}

Some expressions needed in Sec. III. A. 2. are presented here:%
\[
f_{in+AQ}\left(  I_{1},\tau iI_{2},I_{2}^{2}\right)  \equiv\frac{2\tau iI_{2}%
}{\left(  1+I_{1}\right)  ^{2}+\left(  iI_{2}\right)  ^{2}},
\]%
\begin{align*}
f_{sk}^{UEE}\left(  I_{1},I_{2}^{2}\right)   &  \equiv\left(  \frac{1+I_{1}%
}{\left(  1+I_{1}\right)  ^{2}+\left(  iI_{2}\right)  ^{2}}-1\right) \\
&  \times\left(  2\frac{1+I_{1}}{\left(  1+I_{1}\right)  ^{2}+\left(
iI_{2}\right)  ^{2}}-1\right) \\
&  -2\left(  \frac{iI_{2}}{\left(  1+I_{1}\right)  ^{2}+\left(  iI_{2}\right)
^{2}}\right)  ^{2},
\end{align*}
and $f_{L}\left(  I_{1},I_{2}^{2}\right)  \equiv1+\frac{\epsilon_{R}}%
{\epsilon_{F}}-2\frac{\epsilon_{R}}{\epsilon_{F}}\frac{1-I_{1}^{2}-\left(
iI_{2}\right)  ^{2}}{\left(  1+I_{1}\right)  ^{2}+\left(  iI_{2}\right)  ^{2}%
}$.

For the anomalous Hall conductivity, we get $\sigma_{xy}^{b}=\frac{e^{2}}%
{\pi\hbar}\frac{\epsilon_{R}}{\hbar}\tau iI_{2}$, $\sigma_{xy}^{l}=\frac
{e^{2}}{\pi\hbar}\frac{2\epsilon_{R}}{\hbar}f_{in+AQ}\left(  I_{1},\tau
iI_{2},I_{2}^{2}\right)  $ and $\sigma_{xy}^{sk}=-\frac{e^{2}}{\pi\hbar}%
\frac{\epsilon_{R}}{\hbar}\tau\frac{mV_{0}}{\hbar^{2}}f_{sk}^{AHE}\left(
I_{1},I_{2}^{2}\right)  $, where%
\begin{align*}
f_{sk}^{AHE}\left(  I_{1},I_{2}^{2}\right)   &  \equiv\left(  \frac
{1-I_{1}^{2}-\left(  iI_{2}\right)  ^{2}}{\left(  1+I_{1}\right)  ^{2}+\left(
iI_{2}\right)  ^{2}}\right)  ^{2}\\
&  -\left(  \frac{2iI_{2}}{\left(  1+I_{1}\right)  ^{2}+\left(  iI_{2}\right)
^{2}}\right)  ^{2}.
\end{align*}


\begin{thebibliography}{99}
\bibitem {note-SOscattering}In the present paper we do not consider spin-orbit
interaction around the impurities.
\bibitem {note-SKtime}For instance, in isotropic single-Fermi-circle 2D
systems $\tau^{sk}\sim DV_{0}\left(  \tau^{tr}\right)  ^{2}/\tau$, where
$\tau^{tr}$ and $\tau$ are the transport time and lifetime of electrons,
respectively. In the high-mobility 2D electron gas in GaAs heterostructures,
$\tau^{tr}/\tau$ can be larger than 100 if the charged scatterers are placed
far away from the 2D conducting plane. In this case the qualitative analysis
based on the smallness of $DV_{0}$ is practically not reliable.
\bibitem {note-continuation}We note that in the case of strong
disorder-potential the separation of anomalous quantum and skew scattering
contributions in the non-Boltzmann regime with larger impurity density is ambiguous.
\end{thebibliography}
\end{document}